\documentclass[9pt,twocolumn,twoside]{osajnl}

\journal{ol} 

\setboolean{shortarticle}{true}

\title{Effective discrimination of chiral molecules in a cavity}

\author[1,2]{Yi-Hao Kang}
\author[1,2]{Zhi-Cheng Shi}
\author[3]{Jie Song}
\author[1,2]{Yan Xia}

\affil[1]{Fujian Key Laboratory of Quantum Information and Quantum Optics (Fuzhou University), Fuzhou 350116, China}
\affil[2]{Department of Physics, Fuzhou University, Fuzhou 350116, China}
\affil[3]{Department of Physics, Harbin Institute of Technology, Harbin 150001, China}

\affil[*]{Yan Xia: xia-208@163.com}

\begin{abstract}
We present a scheme to realize precise discrimination of chiral molecules in a cavity. Assisted by additional laser pulses, cavity fields can evolve to different coherence states with contrary-sign displacements according to the handedness of molecules.
Consequently, the handedness of molecules can be read out with homodyne measurement on the cavity, and the successful probability
is nearly unity without very strong cavity fields. Numerical results 
show that the scheme is insensitive to errors, noise and decoherence.
Therefore, the scheme may provide helpful perspectives for accurate
discrimination of chiral molecules. 
\end{abstract}

\setboolean{displaycopyright}{true}

\begin{document}

\maketitle

Chiral molecules, composed of enantiomers
being mirror images of each other, have shown many interesting
applications in chemistry, biotechnologies, and pharmaceutics
\cite{BarrettNat509,GalChi24,AmorimIBB115}. Although a pair of
enantiomers share many physical and chemical properties, their own special properties
originated from the handedness show divergent behaviors in chemical
and biological reactions. Therefore, it is necessary to detect and
separate racemic mixtures of enantiomers. However, discrimination of
chiral molecules is usually a difficult task in chemistry
\cite{KnowlesACIEE41}, since chemical techniques, such as
crystallization, derivatization, kinetic resolution, and chiral
chromatography \cite{Book1} are usually time-consuming and
expensive.

Fortunately, researches
\cite{FujimuraCPL306,HokiCP267,GonzalezJCP115,HokiJCP116,JacobJCP137,PattersonNat497} have shown that
electric-dipole interactions between chiral molecules and linearly
polarized light have potential to realize enantioselectivity as some
of transition dipole moments of left ($L$) and right ($R$) handed
enantiomers have different signs. Thus, by using filed-molecule couplings, it
is possible to rapidly detect handedness of molecules based on the
difference of evolutions. From this point, schemes for the
discrimination of chiral molecules with field-molecule couplings are proposed by using techniques
like adiabatic passage \cite{ShapiroPRL84,KralPRL87,KralPRL90} and flat resonant
pulses \cite{LYPRA77}. Interestingly, recent schemes
\cite{VitanovPRL122,WJLPRA100,WJLPRApp13} using the technique
``shortcut to adiabaticity'' \cite{CXPRL105,CampoPRL111,CampoPRL109,MasudaPRA84} 
further make trade-off between the evolution speed and robustness to errors, 
which make discrimination of chiral molecules more effective. However, 
in these previous schemes
\cite{KralPRL87,KralPRL90,LYPRA77,VitanovPRL122,WJLPRA100,WJLPRApp13},
discrimination of enantiomers relies on the population difference,
and requires higher molecule levels and coherence superposition in
the process. As a result, decoherence of molecules, e.g., energy relaxation and
dephasing may cause significant population errors, and pursuing an
acceptable successful probability would impose high requirement on the
coherence time of molecules. On the other hand, because population of a molecule state can only 
ranges between 0 and 1, and is proportional to the successful probability, 
small deviations of pulse areas may also significantly reduce successful probabilities of the discrimination. 
Therefore, if the information of handedness can be stored and read out in some stable ways
instead of using populations of molecules, the discrimination of
chiral molecules can be more accurate in the presence of experimental imperfections. 

In this letter, we propose an effective scheme to discriminate chiral molecules. 
Apart from external driving pulses applied to molecules, we also consider
couplings of molecules with the quantized field in a cavity.
Instead of the population difference, information of handedness of
enantiomers in the scheme can be perfectly acquired from coherence
states of cavity fields with contrary-sign displacements by using 
homodyne measurement
\cite{TorresPRA90,NemotoPRL93,BarrettPRA71,GutierrezPRA99}. 
The error rates of the scheme are described by
complementary error functions of the difference of the displacements. 
When we set proper final displacements of coherence state, the successful 
probabilities are insensitive to deviations of the displacements
caused by systematic errors and random noise.
Thus, compared with schemes using population difference, where successful probabilities 
are proportional to the population difference, the scheme can be more robust against errors and noises. 
Moreover, in the scheme, molecules are almost restricted in their ground
states, thus robustness against energy relaxation and dephasing of
molecules can be enhanced. Numerical results indicates that the
requirement of coherence time in the scheme is much looser than the
previous schemes \cite{WJLPRA100,WJLPRApp13} when energy relaxation,
dephasing and cavity decay are all taken into account. Therefore,
the scheme may provide a feasible way to discriminate chiral
molecules.

\begin{figure}[htbp]
\centering
\hbox{\includegraphics[width=\linewidth]{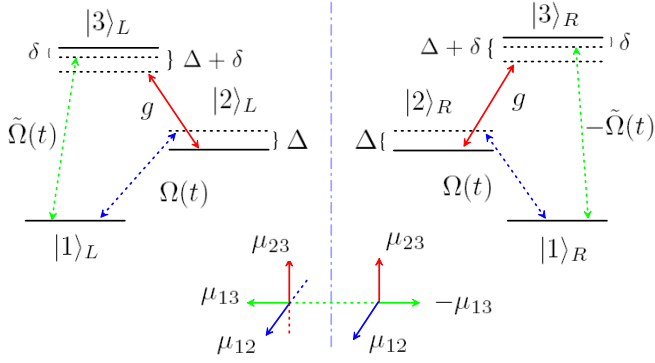}}
\caption{Comparison of the coupling schemes between three discrete
energy states in molecules with left ($L$) and right ($R$)
handedness.}
\label{fig1}
\end{figure}

Let us now describe the physical model in detail. As shown in Fig.~\ref{fig1}, we consider two
enantiomers of chiral molecules with $L$- and $R$-handedness. Both
of $L$- and $R$-handed molecules have three discrete energy states,
denoted by $\{|1\rangle_L,|2\rangle_L,|3\rangle_L\}$ and
$\{|1\rangle_R,|2\rangle_R,|3\rangle_R\}$, respectively. Due to the
difference of handedness, the transition-dipole moment
$\vec{\mu}_{13}$ is in different directions for $L$- and $R$-handed
molecules, while the transition-dipole moment $\vec{\mu}_{12}$
($\vec{\mu}_{23}$) is in the same direction for $L$- and $R$-handed
molecules \cite{VitanovPRL122,WJLPRA100,WJLPRApp13}. We assume the
transition $|1\rangle\leftrightarrow|2\rangle$ is driven by a
classical field $\vec{E}(t)=\varepsilon(t)\cos(\omega
t)\vec{e}_{12}$ with frequency $\omega$, amplitude $\varepsilon(t)$
and polarization direction $\vec{e}_{12}$. Assuming that the
transition frequency of $|1\rangle\leftrightarrow|2\rangle$ is
$\omega_{12}=\omega-\Delta$, the Rabi frequency and the (blue)
detuning of the transition should be
$\Omega(t)=\varepsilon(t)\vec{\mu}_{12}\cdot\vec{e}_{12}$ and
$\Delta$. In addition, the transition
$|2\rangle\leftrightarrow|3\rangle$ couples with a quantized cavity
mode with frequency $\nu$ in a cavity with coupling
strength $g=\sqrt{\frac{\nu}{2\epsilon
V}}\vec{\mu}_{23}\cdot\vec{e}_{23}$, with $\vec{e}_{23}$ being the
unit polarization vector for cavity mode, $\epsilon$ being
dielectric constant of the dielectric of the cavity, and $V$ being the volume
of the cavity. For the transition frequency
$\omega_{23}=\nu+\Delta+\delta$, the (red) detuning of the coupling
is $\Delta+\delta$. Moreover, the transition
$|1\rangle\leftrightarrow|3\rangle$ is also driven by a classical
field
$\vec{\tilde{E}}(t)=\tilde{\varepsilon}(t)\cos(\tilde{\omega}t)\vec{e}_{13}$
with frequency $\tilde{\omega}$, amplitude $\tilde{\varepsilon}(t)$
and polarization direction $\vec{e}_{13}$. For the transition
frequency $\omega_{13}=\tilde{\omega}+\delta$, the (red) detuning of
the transition is $\delta$. However, due to different handedness,
the Rabi frequency is
$\tilde{\Omega}(t)=\tilde{\varepsilon}(t)\vec{\mu}_{13}\cdot\vec{e}_{13}$
for $L$-handed molecule but becomes $-\tilde{\Omega}(t)$ for $R$-hand
molecule. In the rotating-wave-approximation, the total Hamiltonian
of an enantiomer reads
\begin{eqnarray}\label{e1}
H(t)&=&\Omega(t)e^{i\Delta
t}|1\rangle\langle2|+gae^{i(\Delta+\delta)t}|3\rangle\langle2|\cr\cr
&\pm&\tilde{\Omega}(t)e^{-i\delta
t}|1\rangle\langle3|+\mathrm{H.c.},
\end{eqnarray}
with $a$ being the annihilate operator of the cavity mode and
$\pm\tilde{\Omega}(t)$ for $L$- and $R$-handed molecules,
respectively. We consider the conditions $\tilde{\Omega}(t)\sim
g^2/\Delta\ll\delta$, $\Omega(t)\sim g\ll\Delta$ and
$\delta\ll\Delta$. The effective Hamiltonian can be derived via
second-order perturbation theory \cite{JamesCJP85} as
\begin{eqnarray}\label{e2}
H_e(t)&=&H_1(t)+H_2(t)+H_3(t),\cr\cr
H_1(t)&=&-\frac{\Omega^2(t)g^2}{\Delta^2\delta}a^\dag
a|1\rangle\langle1|,\cr\cr
H_2(t)&=&[\frac{\Omega^2(t)}{\Delta}-\frac{\tilde{\Omega}^2(t)}{\delta}]|1\rangle\langle1|,\cr\cr
H_3(t)&=&\mp\frac{\Omega(t)\tilde{\Omega}(t)g}{\Delta\delta}(a^\dag+a)|1\rangle\langle1|.
\end{eqnarray}

Assuming the considered enantiomer is initial in the lowest state
$|1\rangle$, $H_2(t)$ only leads a global phase, thus can be omitted.
In the rotation frame of $\mathcal{R}=\exp[-i\int_0^tH_1(t')dt']$,
the effective Hamiltonian can be simplified as
\begin{eqnarray}\label{e3}
\tilde{H}_e(t)=\mp\Omega_e(t)[e^{-i\varphi(t)}a^\dag+e^{i\varphi(t)}a]|1\rangle\langle1|,
\end{eqnarray}
with $\Omega_e(t)=\Omega(t)\tilde{\Omega}(t)g/\Delta\delta$,
$\varphi(t)=\int_0^tdt'\Omega^2(t')g^2/\Delta^2\delta$, and
$\mp\Omega_e(t)$ for $L$- and $R$-handed molecules, respectively.

Assuming the cavity is initial in the vacuum state, evolution (density
operator) of the system can be calculated as \cite{LuisJPA34}
\begin{eqnarray}\label{e4}
&&\tilde{\rho}_e(t)=|\pm\tilde{\alpha}(t)\rangle_c\langle\pm\tilde{\alpha}(t)|\otimes|1\rangle_e\langle1|,\cr\cr
&&\tilde{\alpha}(t)=\int_0^ti\Omega_e(t')e^{-i\varphi(t')}dt',
\end{eqnarray}
with subscript $c$ and $e$ corresponding to the states of cavity and
enantiomer. In addition, $|\tilde{\alpha}(t)\rangle_c$ is a coherence state of
cavity. Therefore, back to the original frame, the evolution at
final time $t=T$ can be described by
\begin{eqnarray}\label{e5}
\rho_e(T)=|\pm\alpha(T)\rangle_c\langle\pm\alpha(T)|\otimes|1\rangle_e\langle1|,
\end{eqnarray}
with $\alpha(T)=\tilde{\alpha}(T)e^{i\varphi(T)}$. According to the
result, for $L$-handed molecules, the cavity is finally in state
$|\alpha(T)\rangle$, while for $R$-handed molecules, the cavity is
finally in state $|-\alpha(T)\rangle$. When $|\alpha(T)|\gg1$,
$|\langle-\alpha(T)|\alpha(T)\rangle|\sim0$, the difference of
cavity can be easily discriminated by a $X_\phi$-quadrature homodyne
measurement
\cite{TorresPRA90,NemotoPRL93,BarrettPRA71,GutierrezPRA99} with a
probing mode in coherence state $|z\rangle$ with
$\phi=\mathrm{arg}(z)=\mathrm{arg}[\alpha(T)]$. This is equivalent
to the eigenprojection $|x,\phi\rangle\langle x,\phi|$ of operator
$X_\phi=e^{i\phi}a^\dag+e^{-i\phi}a$. Considering the probability
distribution $f_\pm(x,\phi)=|\langle
x,\phi|\pm\alpha(T)\rangle|^2=(2/\pi)^{1/2}\exp[-2(x\pm|\alpha|)^2]$
\cite{TorresPRA90,GutierrezPRA99}, if we consider result of homodyne
measurement result with $x>0$ ($x<0$) for the cavity in state
$|\alpha(T)\rangle$ ($|-\alpha(T)\rangle$), the error probability is
$P_e=\int^{+\infty}_0f_-(x,\phi)dx=\mathrm{erfc}[\sqrt{2}|\alpha(T)|]/2$,
which becomes less than $3.16712\times10^{-5}$ when $|\alpha(T)|>2$.
Therefore, the scheme can give near deterministic discrimination of
chiral molecules.

\begin{figure}[htbp]
\centering
\hbox{\includegraphics[width=\linewidth]{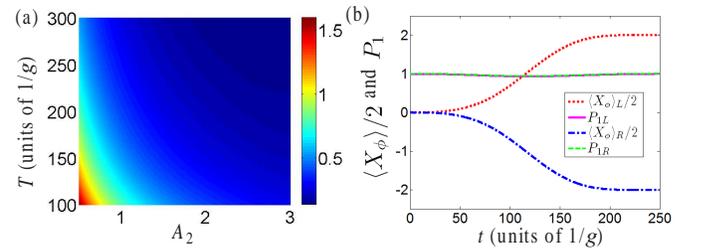}}
\caption{(a)$A_1=A_2g/[\Delta\sin(\pi\mu)]$ versus $A_2$ and $T$. (b) $\langle
X_\phi\rangle/2$ and $P_1$ versus $t$ with $L$-handed molecule
(red-dotted line: $\langle X_\phi\rangle_L/2$; magenta-solid line:
$P_{1L}$) and $R$-handed molecule (blue-dashed-and-dotted line:
$\langle X_\phi\rangle_R/2$; green-dashed line:
$P_{1R}$).}
\label{fig2}
\end{figure}

We now using numerical simulations to show the
performance of the scheme. Using the coupling strength $g$ as a
reference value, we select control functions as
$\tilde{\Omega}(t)=A_1g\sin(\pi t/T)$ and $\Omega(t)=A_2g\sin(\pi
t/T)$, with two dimensionless parameters $A_1$ and $A_2$. In this
way, we obtain
$\alpha(T)=A_1A_2g^3T\exp\{i[\pi(1/2-\mu)+\varphi(T)]\}\sin(\pi\mu)/(2\pi\mu\Delta\delta)$,
$\mu=A_2^2g^4T/(4\pi\Delta^2\delta)$, and
$\varphi(T)=A_2^2g^4T/(2\Delta^2\delta)$. Accordingly, the angle
should be $\phi=\pi(1/2-\mu)+\varphi(T)$ in the $X_\phi$-quadrature
homodyne measurement. In addition, $A_1\geq
A_2g/[\Delta\sin(\pi\mu)]$ should be satisfied to make $|\alpha(T)|\geq2$.
To examine the relations between parameters $A_1$, $A_2$ and $T$, we
first set $\Delta=20g$ and $\delta=20g^2/\Delta$ for conditions
$\Delta\gg g$ and $\delta\gg g^2/\Delta$, and we plot
$A_1=A_2g/[\Delta\sin(\pi\mu)]$ versus $A_2$ and $T$ in
Fig.~\ref{fig2}(a).
As shown in Fig.~\ref{fig2}(a), when $\Delta\gg g$ and $\delta\gg
g^2/\Delta$, $A_1$ decreases when $A_2$ and $T$ increase. Since
$\tilde{\Omega}(t)\sim g^2/\Delta$ should be satisfied, we consider a
proper value $A_1=0.15$ at point $(A_2,T)=(2.5,250/g)$. With the
selections of parameters, we plot the variation of the average value
$\langle X_\phi\rangle/2=\mathrm{Tr}[X_\phi\rho(t)]/2$ versus $t$ in
Fig.~\ref{fig2}(b). According to Fig.~\ref{fig2}(b), $\langle X_\phi\rangle/2$
varies from 0 to $\pm2.068$ when the considered molecules are $L$- and
$R$-handed, respectively. Noticing that $\langle
X_\phi\rangle=2|\alpha(T)|$ is satisfied when
$\phi=\mathrm{arg}[\alpha(T)]$, the numerical result accords with the
theory analysis based on the effective Hamiltonian. Considering a
reported coupling strength $g=10$MHz \cite{LWPRA76} for molecules in cavity, the total interaction time required is $T=25\mu$s. Thus, $L$- and $R$-handed molecules can be fast discriminated via homodyne measurement with error rates below $3.16712\times10^{-5}$. Moreover, we also examine the population
$P_1$ of state $|1\rangle$ in the evolution. Seen from the
magenta-solid line and the green-dashed line, populations of both $L$- and
$R$-handed molecules almost keep at 1 during the evolution. This
also proves the validity of the theory analysis.

We now demonstrate the performance of the scheme in the presence of experimental imperfections. First, due to operational and instrumental imperfection, 
there may exist systematic errors in the Rabi frequencies and coupling strengths. For the scheme, we assume
systematic errors as
$\tilde{\Omega}(t)\rightarrow(1+\eta_1)\tilde{\Omega}(t)$,
$\Omega(t)\rightarrow(1+\eta_2)\Omega(t)$ and
$g\rightarrow(1+\eta_3)g$ with $\eta_j$ $(j=1,2,3)$ being the faulty coefficient. We plot $D(X/2)=(\langle X_\phi\rangle_L-\langle
X_\phi\rangle_R)/2$ versus $\eta_j$ in Fig.~3(a). In Fig.~3(a), with the increase of $\eta_1$ from 0 to 0.1, $D(X/2)$ also increases. Theoretically, according to the effective Hamiltonian, systematic
error of $\tilde{\Omega}(t)$ only influences the displacement $\alpha(T)$, consequently $|\alpha(T)|$ increases when $\eta_1>0$.
However, systematic errors of $\Omega(t)$ and $g$ influence both displacement $\alpha(T)$ and angle $\varphi(T)$, thus both increase
and decrease of $\eta_2$ and $\eta_3$ may reduce $D(X/2)$ due to the mismatching of angles. According to the results, when $\eta_j\in[-0.1,0.1]$, $D(X/2)$ remains higher than 3.77. Therefore, the error rate of homodyne measurement is still below $8.162\times10^{-5}$. In addition, we examine the systematic errors of 
detunings as $\Delta\rightarrow(1+\eta_1')\Delta$ and
$\delta\rightarrow(1+\eta_2')\delta$. $D(X/2)$ versus $\eta_1'$ and $\eta_2'$ is shown in Fig.~3(b). The result indicates that $D(X/2)$ tends to
decrease with the increase of $\eta_1'$ and $\eta_2'$ when
$\eta_1',\eta_2'\in[-0.1,0.1]$. This is because $\Omega_e(t)$ decreases when
$\Delta$ and $\delta$ increase. The worst case appears at
$(\eta_1',\eta_2')=(0.1,0.1)$, which gives $D(X/2)=3.601$. In this case, the error rate of homodyne measurement is only $P_e=1.5911\times10^{-4}$. Therefore, the scheme provides nice fault-tolerance for the discrimination of chiral molecules.

Apart from systematic errors, random noise due to fluctuation of experimental environment also disturb evolutions of physical systems. As the additive white Gaussian noise (AWGN) is a nice model
to describe random noise, we here consider the influence of AWGN in the pulses. Rabi frequency $Y(t)\in\{\tilde{\Omega}(t),\Omega(t)\}$
under the influence of the AWGN is
$Y_{N}(t)=Y(t)+\mathrm{awgn}(Y(t),R_N)$, where
$\mathrm{awgn}(Y(t),R_N)$ is a function generating AWGN with signal-to-noise ratio (SNR) $R_N$ for pulse, and $Y_N(t)$ is the pulse involves AWGN. To estimate the influence of AWGN in average,
we perform fifty simulations with SNR $R_N=10$, and the results are shown in Fig.~3(c). According to the results, the worst result in
the fifty simulations is $D(X/2)=3.215$, corresponding to error rate $P_e=6.5222\times10^{-4}$. Therefore, the scheme also holds good robustness against random noise.

\begin{figure}[htbp]
\centering
\hbox{\includegraphics[width=0.93\linewidth]{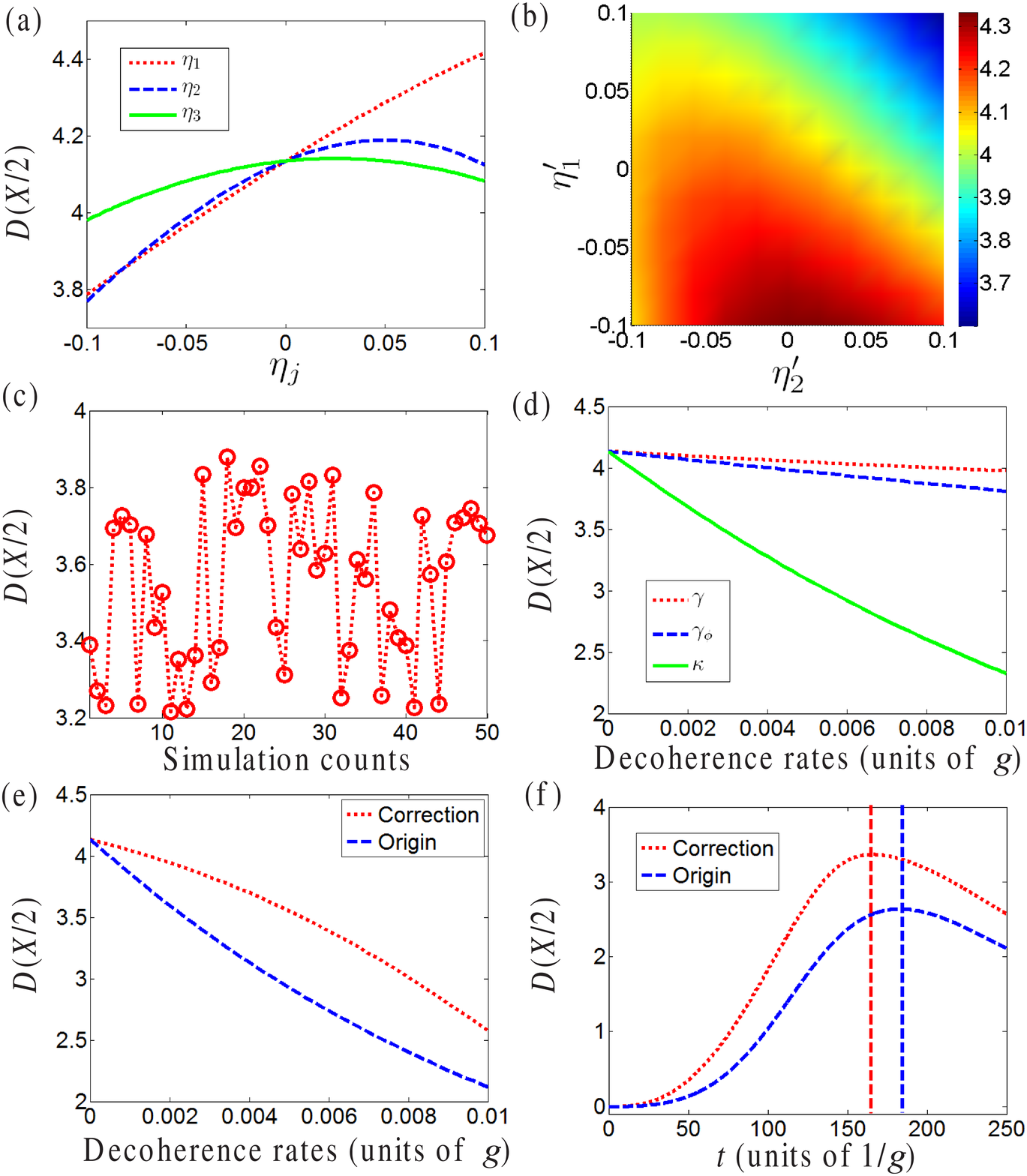}}
\caption{(a)
$D(X/2)$ versus $\eta_j$. (b) $D(X/2)$ versus $\eta_1'$ and
$\eta_2'$. (c) $D(X/2)$ versus simulation counts under AWGN with SNR
$R_N=10$. (d) $D(X/2)$ versus decoherence rates $\gamma$,
$\gamma_\phi$ and $\kappa$. (e) $D(X/2)$ versus decoherence rate
($\gamma=\gamma_\phi=\kappa$) with corrected pulses and original
pules. (f) $D(X/2)$ versus $t$ with corrected pulses and original
pules ($\gamma=\gamma_\phi=\kappa=0.01g$).}
\label{fig3}
\end{figure}

Finally, let us consider the influence of decoherence to the
operation. For molecules, the energy relaxation and dephasing are
two disturbing factor destroying the coherence. On the other hand,
the cavity decay weakens intensity of the cavity field, which may also reduce the successful probability of homodyne measurement. In the
presence of these decoherence factors, the evolution is governed by
a master equation as
\begin{eqnarray}\label{e6}
\dot{\rho}(t)&=&-i[H(t),\rho(t)]+\frac{\kappa}{2}\mathcal{L}[a]\rho(t)\cr\cr
&+&\sum\limits_{\iota=1}^{2}\sum\limits_{\iota'=\iota+1}^{3}\{\frac{\gamma}{2}\mathcal{L}[\sigma^-_{\iota\iota'}]\rho(t)
+\frac{\gamma_{\phi}}{2}\mathcal{L}[\sigma^z_{\iota\iota'}]\rho(t)\},
\end{eqnarray}
with $\mathcal{L}[o]\rho(t)=2o\rho(t)o^\dag-o^\dag o\rho(t)-\rho(t)o^\dag o$,
cavity decay rate $\kappa$, energy relaxation rate $\gamma$,
dephasing rate $\gamma_{\phi}$,
$\sigma^-_{\iota\iota'}=|\iota\rangle\langle\iota'|$, and
$\sigma^z_{\iota\iota'}=(|\iota'\rangle\langle\iota'|-|\iota\rangle\langle\iota|)/2$.
We plot $D(X/2)$ versus different decoherence rates in Fig.~3(d).
The results show that when $\kappa=0.01g$, $\gamma=0.01g$, and
$\gamma_{\phi}=0.01g$, we obtain $D(X/2)=2.2326$, $D(X/2)=3.978$,
and $D(X/2)=3.812$, corresponding to error rates $P_e=0.012$,
$P_e=3.47\times10^{-5}$, and $P_e=6.89\times10^{-5}$, respectively.
For $g=10$MHz \cite{LWPRA76}, the docoherence rate/coherence time is
about 100kHz/10$\mu$s. In the previous schemes
\cite{WJLPRA100,WJLPRApp13} for discrimination of chiral molecules
by using difference of population, due to the effect of energy
relaxation, long lifetime of higher molecule levels
about $\tau\sim200\mu$s-$300\mu$s ($\gamma\sim3.3$kHz-5kHz) is required to get
acceptable successful probabilities ($\sim0.99$). However, it is hard for a
higher molecule level to possess such long lifetime for various
types of molecules. Moreover, the effect of dephasing is also not
negligible in practice. When the dephasing is taken into account,
the successful probability of schemes \cite{WJLPRA100,WJLPRApp13} may be further reduced. In the
current scheme, since the molecule level is almost restricted in the
ground state, the evolution is insensitive to both energy relaxation and
dephasing, thus greatly relax the requirement of coherence time.
Although cavity decay caused more influence compared the other two
factors, the error rate $P_e=0.012$ is still acceptable to
discriminate chiral molecule. In fact, errors caused by cavity decay can be further
reduced. When cavity decay is considered, the evolution can be
analyzed with superoperators as
$\rho(t+dt)=e^{\mathcal{L}[a]dt}\mathcal{U}(t,dt)\rho(t)$, with
$\mathcal{U}(t,dt)\rho(t)=e^{-iH_e(t)dt}\rho(t)e^{iH_e(t)dt}$
\cite{ZSBPRA91}. Noticing the result
\begin{eqnarray}\label{e7}
&&e^{\mathcal{L}[a]dt}|\alpha\rangle\langle\alpha|
=(1+\mathcal{L}[a]dt)|\alpha\rangle\langle\alpha|+\mathcal{O}(dt^2)\cr\cr
&&=|\alpha(1-\kappa dt/2)\rangle\langle\alpha(1-\kappa
dt/2)|+\mathcal{O}(dt^2),
\end{eqnarray}
the evolution can be calculated as
\begin{eqnarray}\label{e8}
\bar{\rho}(T)=|\pm\bar{\alpha}(T)\rangle_c\langle\pm\bar{\alpha}(T)|\otimes|1\rangle_e\langle1|,\cr\cr
\bar{\alpha}(T)=e^{i\varphi(T)-\kappa
T/2}\int_0^Ti\Omega_e(t)e^{\kappa t/2-i\varphi(t)}dt.
\end{eqnarray}

Although cavity decay is hard to avoid, the decay rate may be predetermined via experiments. When decay rate is known, we can make
correction of pulse as
$\tilde{\Omega}(t)\rightarrow\tilde{\Omega}(t)\exp[\kappa(T-t)/2]$
to make $\bar{\alpha}(T)=\alpha(T)$. As shown in Fig.~3(e), $D(X/2)$
versus decoherence rate ($\gamma=\gamma_\phi=\kappa$) is plotted
with corrected pulses and original pulses. The result indicate that
when the corrected pulses are applied, the discrimination is
improved a lot compared with that using original pulses. For
$\gamma=\gamma_\phi=\kappa=0.01g$, we still have $D(X/2)=2.575$,
corresponding to error rate $P_e=0.005$. In addition, we can also
reduce error rates by measuring cavity fields at a time $T'$ before $T$ since the maximal value of $D(X/2)$ does not appear at $T$ when decoherence is included. As an example, we plot $D(X/2)$ versus $t/T$ with
$\gamma=\gamma_\phi=\kappa=0.01g$ in Fig.~\ref{fig3}(f), where we can find that with
the correction, the maximum $D(X/2)=3.367$ ($P_e=3.8\times10^{-4}$)
appears at $T'=165/g$, and without the correction, the maximum $D(X/2)=2.639$ ($P_e=0.004$) appears at $T'=183/g$. By measuring cavity fields at $t=T'$, the
successful probabilities can be significantly improved. Furthermore,
recent schemes for intensifying coupling strengths \cite{QWPRL120} or
improving quality factors \cite{AcharyyaPRapp12} make it possible to
enhance the ratio $g/\kappa$ in a cavity. These techniques
can make the scheme works more efficiently in a real scenario.

As an example for the implementation of the scheme, the rotational states $|0_{00}\rangle$, $|1_{11}\rangle$ and
$|1_{10}\rangle$ of 1,2-propanediol molecules with $|J_{k_{-1},k_1}\rangle$ being the rotational state with quantum numbers of the limiting prolate and
oblate symmetric top $k_{-1}$
and $k_1$, can be considered as levels $|1\rangle$, $|2\rangle$ and $|3\rangle$, respectively \cite{PattersonNat497}. The
transition frequencies between these three states are
$\omega_{12}=11363$MHz, $\omega_{13}=12212$MHz, and
$\omega_{2,3}=849$MHz in this model, and the frequencies of pulses $\Omega(t)$ and $\tilde{\Omega}(t)$ should be $\omega=11563$MHz and $\tilde{\omega}=12202$MHz. In addition, the frequency of the cavity mode is $\nu=639$MHz.

In conclusion, we have proposed an effective
scheme to discriminate chiral molecule with the help of a
cavity. We showed that enantiomers possessing different handedness
produce coherence states of cavity mode with contrary-sign
displacements $\pm\alpha(T)$, consequently information about the
handedness can be read out by homodyne measurement
\cite{TorresPRA90,NemotoPRL93,BarrettPRA71,GutierrezPRA99} with very
slight error rates. The scheme does not require very strong
coherence state. When $|\alpha(T)|=2$ (average photon number
$\bar{n}=4$), the error rate is only $3.16712\times10^{-5}$ in
theory. Moreover, the scheme is robust against deviations of
displacements caused by systematic errors and random noise. Because
when $|\alpha(T)|$ is big enough, deviations of displacements only
induce tiny change of error rates. Compared with the scheme
\cite{WJLPRA100,WJLPRApp13} based on difference of populations, the
scheme provide a fault-tolerance approach for the discrimination
since populations of molecules are more sensitive to errors of
pulses. Furthermore, as molecules are almost restricted in the
ground states, the scheme also holds better robustness against
energy relaxation and dephasing of molecules. With synthetical
consideration of energy relaxation, dephasing and cavity decay, the
scheme can still produce acceptable successful probability with
coherence time only about $10\mu$s. Therefore, the scheme may be
useful to realize accurate discrimination of chiral molecule in the
presence of experimental imperfections.

\noindent\textbf{Funding.} National Natural Science Foundation of China (11805036)

\noindent\textbf{Disclosures.} The authors declare no conflicts of interest.

\bibliographyfullrefs{sample}

\end{document}